**Young children's anthropomorphism of an AI chatbot: Brain activation and the role of parent co-presence**

Pilyoung Kim[1,2], Jenna H. Chin[1], Yun Xie[1], Nolan Brady[3], Tom Yeh[3], Sujin Yang[2]

[1]Department of Psychology, University of Denver, 2155 S. Race St, Denver, CO, 80208, USA
[2]Department of Psychology, Ewha Womans University, 52 Ewhayeodae-gil, Seodaemun-gu, Seoul, 03760, South Korea
[3]Department of Computer Science, University of Colorado Boulder, 552 UCB, Boulder, 80309, CO, USA

Address correspondence to
Pilyoung Kim, Ph.D.
Department of Psychology
University of Denver
2155 South Race Street
Denver, CO 80208-3500
**Email**: pilyoung.kim@du.edu

**Abstract**

Artificial Intelligence (AI) chatbots powered by a large language model (LLM) are entering young children's learning and play, yet little is known about how young children construe these agents or how such construals relate to engagement. We examined anthropomorphism of a social AI chatbot during collaborative storytelling and asked how children's attributions related to their behavior and prefrontal activation. Children at ages 5-6 (N = 23) completed three storytelling sessions: interacting with (1) an AI chatbot only, (2) a parent only, and (3) the AI and a parent together. After the sessions, children completed an interview assessing anthropomorphism toward both the AI chatbot and the parent. Behavioral engagement was indexed by the conversational turn count (CTC) ratio, and concurrent fNIRS measured oxygenated hemoglobin in bilateral vmPFC and dmPFC regions. Children reported higher anthropomorphism for parents than for the AI chatbot overall, although AI ratings were relatively high for perceptive abilities and epistemic states. Anthropomorphism was not associated with CTC. In the right dmPFC, higher perceptive scores were associated with greater activation during the AI-only condition and with lower activation during the AI+Parent condition. Exploratory analyses indicated that higher dmPFC activation during the AI-only condition correlated with higher end-of-session "scared" mood ratings. Findings suggest that stronger perceptive anthropomorphism can be associated with greater brain activation related to interpreting the AI's mental states, whereas parent co-presence may help some children interpret and regulate novel AI interactions. These results may have design implications for encouraging parent–AI co-use in early childhood.

**Keywords**

Anthropomorphism; Child–AI interaction; Conversational agents; Parent co-presence; Prefrontal cortex; fNIRS

## 1. Introduction

Generative artificial intelligence (AI) has improved rapidly in its ability to hold natural, multi-turn conversations. Conversational agents powered by generative models, such as ChatGPT and Gemini, demonstrate increasingly sophisticated social and emotional responsiveness, often matching or even exceeding human capacities in certain contexts (Ayers et al., 2023; Kaffee et al., 2025). This rapid advancement has generated widespread enthusiasm for the educational potential of generative AI, particularly its promise for personalized learning and creativity support (Elgarf et al., 2024; Giannakos et al., 2025; Su, 2025). Alongside this optimism, there is growing interest in leveraging generative AI to support early childhood development (Kanders et al., 2024; Su, 2025; Yang et al., 2024). Educational and entertainment applications targeting children in this age range are now commonplace, with nearly one-third of parents of children ages 0–8 reporting that their child had used AI for school-related learning in early 2025 (Mann et al., 2025).

At the same time, closer and more frequent interactions between children and AI raise several concerns. Children may place too much trust in an AI partner or rely on it even when it is not accurate (Andries & Robertson, 2023; Solyst et al., 2024). It can be especially difficult for young children to distinguish between simulated abilities and genuine human abilities (Andries & Robertson, 2023). A recent U.S. survey with teens shows wide use of AI companions and reports that one-third of AI users find conversations with AI as satisfying as or more satisfying than those with friends, suggesting a strong social attraction in older youth (Common Sense Media, 2025). Similarly, young children ages 5–7 attributed human-like qualities, such as having thoughts and feelings, to smart speakers and considered them part of the family dynamic (Garg & Sengupta, 2020).

In this context, anthropomorphism plays a central role in shaping how children perceive and relate to AI chatbots. Anthropomorphism refers to the tendency to attribute humanlike qualities such as feelings, intentions, and knowledge to non-human entities (Epley et al., 2007). Classic theory explains anthropomorphism via elicited agent knowledge, effectance motivation, and sociality motivation (Epley et al., 2007). The more humanlike the agent appears, the stronger these attributions tend to be (Waytz, Cacioppo, et al., 2010). Developmentally, younger children (e.g., ages 5–6) often ascribe richer mental states to robots and voice agents than older children and adolescents do (Girouard-Hallam et al., 2021; Manzi et al., 2020; Su, 2025; van Straten et al., 2020). Among young children, anthropomorphism is multidimensional, encompassing perceptive (e.g., seeing, hearing), emotive (e.g., happiness, sadness), imaginative (e.g., pretending, joking), intentions/desires (e.g., wanting, preferring), and epistemic (e.g., understanding, learning) attributions (Manzi et al., 2020). These domains specify the kinds of cues—such as sensory displays, affective expressivity, goal-directed actions, and evidence of knowledge updating—that are most likely to elicit mind perception in young children. Anthropomorphism can be beneficial (e.g., engaging, empathic interfaces that support learning or therapy) yet also a vulnerability when children form close attachments, overestimate competence, or generalize uncritically from AI guidance (Kory-Westlund & Breazeal, 2019a, 2019b).

Given this mix of benefits and risks, it is important to understand how anthropomorphism shapes children's behavior and feelings during interaction, and how parents might buffer possible risks. In early childhood, novelty and ambiguity about an AI's agency can be high (Xu et al., 2024). Work on joint media engagement shows that when parents are present, they can scaffold learning and help regulate emotions in new media contexts (Ewin et al., 2021; Manzi et al., 2020; Stevens & Takeuchi, 2011; Wood et al., 2016). Thus, co-presence may temper over-attribution, help repair misunderstandings, and keep children engaged. However, few studies have examined children's interactions with AI chatbots under conditions of parent co-presence.

There are limitations among young children in articulating their feelings and thoughts verbally. Thus, to complement self-report and behavior, we index neural engagement during the interaction. We use functional near-infrared spectroscopy (fNIRS) while children participate in conversational storytelling with an AI chatbot, with a parent, and with both together. fNIRS is suitable for young children because it is portable, more tolerant of movement than many neuroimaging methods, and child-friendly in naturalistic tasks (Doherty et al., 2023; Pinti et al., 2020). We focus on the prefrontal cortex because prior work links it to cognitive control, social processing, and emotion regulation (Martin & Ochsner, 2016; Miller & Cohen, 2001; Moriguchi & Hiraki, 2013). Within the prefrontal cortex, the medial sector is especially relevant to anthropomorphic responses. The ventromedial prefrontal cortex (vmPFC) contributes to social valuation and affective empathy, while the dorsomedial prefrontal cortex (dmPFC) supports cognitive perspective taking and broader mentalizing operations (Healey & Grossman, 2018; Shamay-Tsoory, 2011). Thus, medial prefrontal responses are relevant to judgments about the minds of artificial agents, and stronger humanlike attributions may recruit greater engagement of this system (Waytz, Morewedge, et al., 2010). A study with adults reports greater dmPFC activation for more humanlike social robots (Krach et al., 2008).

The current study uses a collaborative storytelling paradigm to examine how anthropomorphism shapes young children's engagement with an AI chatbot. We selected collaborative storytelling because it is a common activity for children at this age with parents and other adults, and AI technologies are increasingly available to support this activity in young children (Elgarf et al., 2024; Fan et al., 2024; Fan et al., 2025; Garg et al., 2022; Han & Cai, 2023; Zhang et al., 2022). Children interacted in three contexts: with the AI chatbot alone, with a parent alone, and with both together. During the interactions, child brain activation in the prefrontal cortex was recorded using fNIRS. This design links self-reported anthropomorphism, behavioral engagement (e.g., conversational turns), and brain engagement.

Using this design, we aimed to answer three questions. First, to what extent do young children anthropomorphize an AI chatbot during collaborative storytelling? Second, how are individual differences in anthropomorphism associated with behavioral and brain engagement, specifically, conversational turn-taking and prefrontal activation in vmPFC and dmPFC, during AI versus parent interactions? Third, does parent co-presence modulate children's affect and engagement during AI interaction? Based on prior literature, we hypothesized that anthropomorphism scores would be higher for parents than for the AI chatbot overall, although children would still, on average, exhibit strong anthropomorphic tendencies toward the AI chatbot. Within the AI condition, we expected that higher anthropomorphism would be associated with greater engagement, indexed by more conversational turns and a preference for the AI as the interaction partner, and with stronger vmPFC and dmPFC activation. We also expected that parent co-presence would attenuate brain responses reflecting cognitive and emotional control demands while maintaining or enhancing behavioral engagement.

## 2. Methods

### *2.1.Participants*

Eligibility criteria for the child participants included being aged 5-6, English-speaking, and having no neurological disorders. A total of 23 children and their parents enrolled in the study. The participants were recruited from the participant pool in the researchers' department. The families were initially contacted by phone to assess their interest in the study. Child participants were 42.9% male, aged 5.8 years ± 0.58 years,100% White/Caucasian, 9.5% Hispanic or Latino, 4.8% American Indian/Alaska Native. The parent participants were 81.0% female, 100% White/Caucasian, 4.8% Hispanic or Latino (Table 1).

After reviewing the completion of the data and data quality, a total of 18 children were included to the fNIRS analysis after 5 children were removed due to motion artifact and missing/excluded data.

This study was preregistered on OSF Registries (OSF Preregistration schema; DOI: 10.17605/OSF.IO/NBRCD; https://osf.io/nbrcd; registered September 22, 2025).

### *2.2.Procedure*

A parent and their child visited a university-affiliated research lab. Upon arrival, a trained researcher reviewed the consent form with the parent, who subsequently provided written informed consent. All procedures were approved by the appropriate Institutional Review Board at the affiliated university.

The session followed in the following structured sequence (Figure 1):

1. Introduction to AI Chatbot
   Children and their parents were introduced to an AI-based character, Fluffo, developed using ChatGPT. Fluffo initiated a brief introductory exchange by greeting the child and asking for their first name. In turn, children and parents were prompted to engage Fluffo with questions (e.g., about its favorite food), which Fluffo reciprocated with similar queries. A visual light cue indicated when the speech-to-text system was active, signaling participants to speak.
2. Creative Storytelling Task
   Participants were introduced to the central storytelling activity, designed to elicit co-narration across conditions. The researcher began by reading an example story aloud to model the expected structure (i.e., beginning, middle, end). Parents and children were then given five minutes to co-create a drawing that would serve as a narrative prompt. During this time, they were encouraged to discuss key story elements such as setting and protagonist. Parents were instructed to provide scaffolding as needed, offering suggestions or ideas to support children who struggled to initiate.

3. fNIRS Cap Fitting and Calibration
   Next, children were introduced to functional near-infrared spectroscopy (fNIRS), a non-invasive neuroimaging method used to assess cortical activity via light-based measurements of blood flow. After explaining the procedure, the researcher measured each child’s head circumference to determine appropriate cap size. Optodes were affixed to the selected cap, which was then fitted to the child's head.
4. Interaction Conditions
   The primary experimental tasks followed, consisting of three conditions. The order of the two conditions, (1) child–AI Chatbot and (2) child–parent storytelling, was counterbalanced across participants, with a third, exploratory condition (AI Chatbot–parent–child triad) consistently administered last (Figure 1 and 2). Each interaction lasted up to 10 minutes. Between the conditions, children had an additional 2-minute window to elaborate on their original drawing. In the AI Chatbot condition, Fluffo greeted the child by name and prompted them to narrate their story. All AI-generated content was reviewed by the researcher prior to text-to-speech conversion to ensure appropriateness. In the triadic condition, children interacted with both their parent and Fluffo, with the parent seated nearby. All sessions were video-recorded and audio-recorded for subsequent transcription and coding.
5. Post-Interaction Interview and Questionnaires
   Following the experimental conditions, a semi-structured interview was conducted with the child to assess their experiences. Parents completed a questionnaire assessing demographic variables and the child’s familiarity with and access to digital technologies.
6. Compensation and Closing
   Families received gift cards as compensation for their time. Each child was also given a personalized picture book generated by AI Chatbot, incorporating images from the storytelling task.

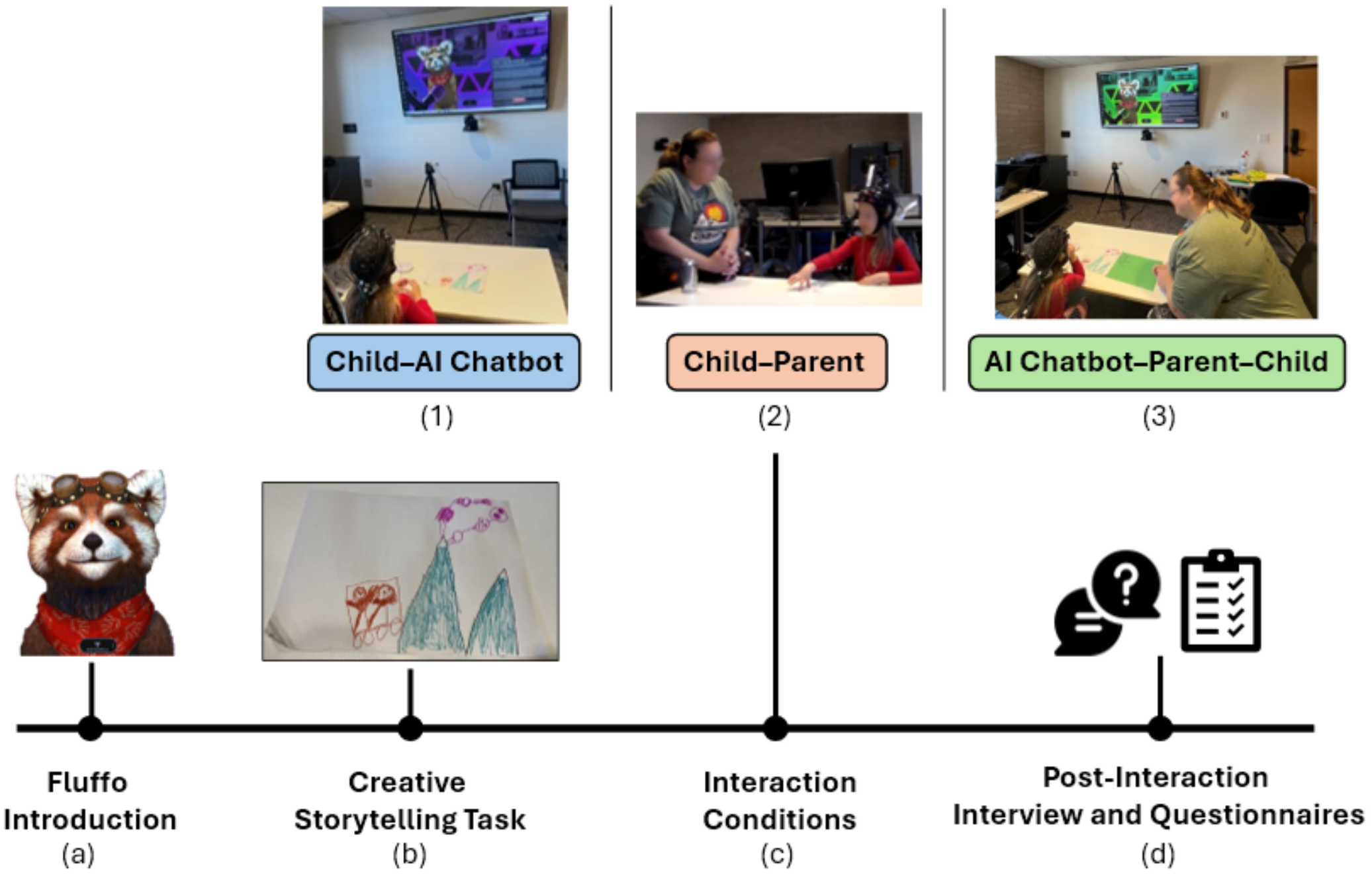

**Figure 1.** Overview of the study procedure. The session included four main components: (a) introduction to the AI Chatbot *Fluffo*; (b) creative storytelling task; (c) interaction conditions: (1) Child–AI Chatbot, (2) Child–Parent, and (3) AI Chatbot–Parent–Child; and (d) post-interaction interview and questionnaires assessing participants' experiences.

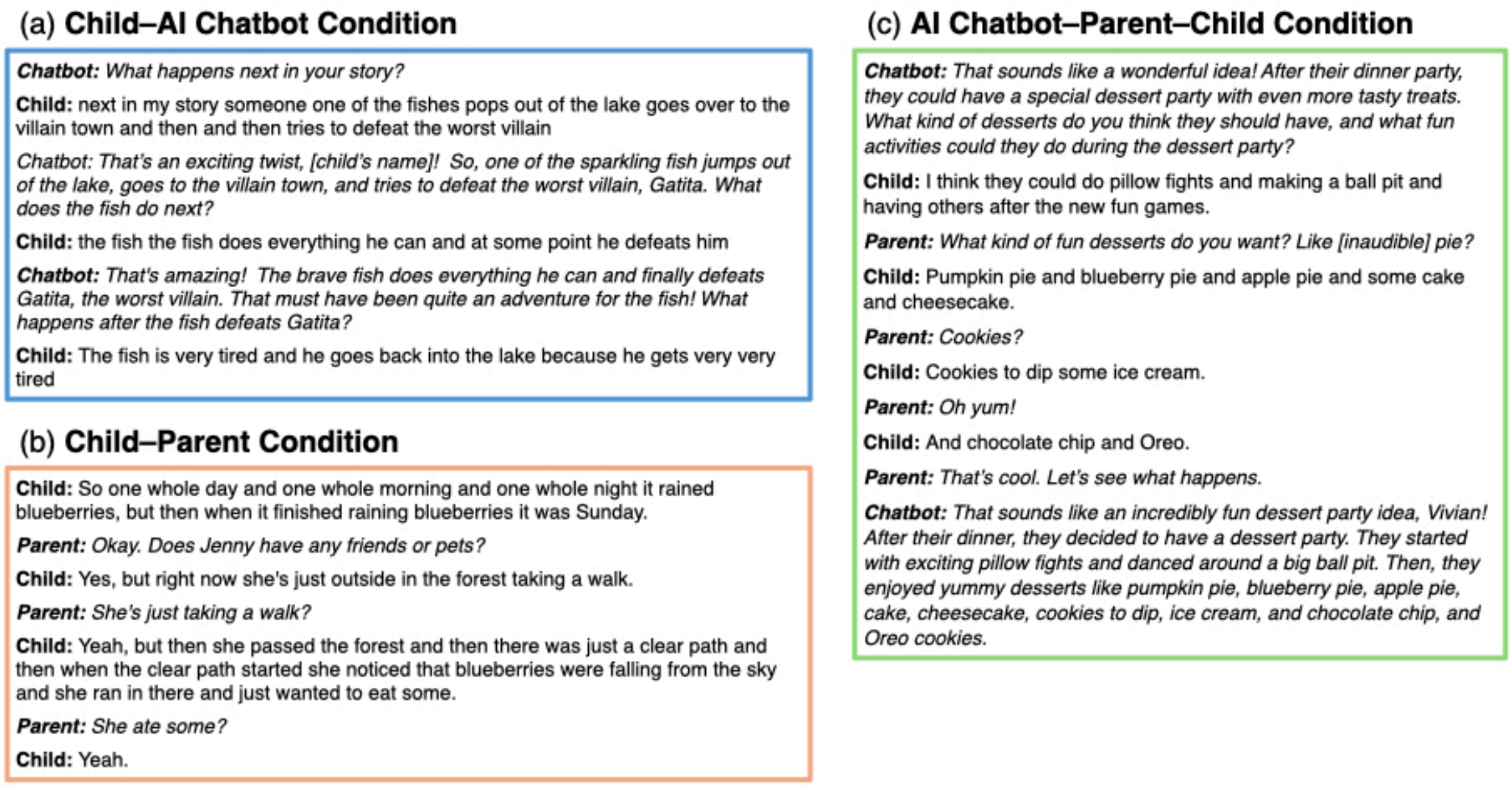

**Figure 2.** Example participant storytelling transcripts from each of the three interaction conditions: (a) Child–AI Chatbot, (b) Child–Parent, and (c) AI Chatbot–Parent–Child

*2.3. Surveys and Interviews*

Survey and interview measures were administered by trained researchers. Parents reported demographics and their child's exposure to technology. The child interview with a researcher included children's attribution of mental states and subjective experience of the storytelling conditions.

**(1)** Demographic Information
Parents provided basic demographic data about themselves and their child, including age, gender, and education level.

**(2)** Child Semi-structured Interview
Immediately after the final session, children reported their current mood (e.g., sad, scared, angry, happy). The interview then assessed their preferences for each interaction condition and which conversational partner they would choose for a subsequent session.

**(3)** Attribution of Mental States Questionnaire
Children's attributions of psychological capacities to social and artificial agents were assessed with the 25-item Attribution of Mental States (AMS) questionnaire (Di Dio et al., 2020; Manzi et al., 2020). Administered as an interview appropriate for young children (used with ages 3–9), the AMS presents yes/no items alongside an image of the target agent (e.g., human, robot) and probes five domains: perceptive abilities (e.g., seeing, hearing), emotive states (e.g., happiness, sadness), intentions/desires (e.g., wanting, preferring), imaginative abilities (e.g., pretending, joking), and epistemic states (e.g., understanding, learning). For each subscale, a composite score from 0–5 was computed (one point per endorsed item), with higher scores indicating greater attribution of mental states to the agent. The measure has been validated for use with children at age 4-6 (Di Dio et al., 2020; Manzi et al., 2020). Owing to substantial missingness in the intentions/desires subscale (>50%; 12/23 cases), this subscale was excluded from subsequent analyses.

**(4)** Child's Exposure to Technology
Parents reported their child's prior experiences with digital technologies. This included frequency and type of interaction with AI-based agents (e.g., ChatGPT), smart speakers (e.g., Siri, Alexa), interactive robotic toys, and applications accessed via smartphones and tablets. One item included a parallel question assessing parental experience with ChatGPT (Supplementary Materials).

*2.4. AI Chatbot System Configuration and Interaction Design*

The AI-based storytelling system was powered by GPT-4, with prompts iteratively developed through pilot sessions involving three children aged 5–6 years. Prompts were crafted to elicit socially responsive, age-appropriate dialogue conducive to co-narration. To enhance the interactivity and appeal of the interface, the Animaze Desktop platform (version 1.26.12546) was used to render a child-friendly avatar. The avatar selected for this study—a red panda named "Fluffo"—featured expressive facial animations and high-resolution graphics. Fluffo was introduced as a fictional character residing in the nearby mountains and was designed to establish rapport by greeting each child by name and providing socially contingent responses.

Children's speech inputs were captured using a Google Chrome speech-to-text browser extension integrated into the ChatGPT interface. Prompts within the system were designed to scaffold the storytelling task through open-ended inquiries, personalized feedback, and story restatements incorporating child input.

### 2.5. *Behavioral Measure of Child Engagement*

To assess children's behavioral engagement during storytelling, conversational turn count (CTC) was analyzed across interaction conditions. These measures provided quantitative indices of the child's verbal participation and dyadic or triadic engagement with each partner (AI Chatbot or/and parent).

Audio recordings of parent-child storytelling sessions were preprocessed in Audacity® (version 3.4.2). Preprocessing steps included amplification and segmentation to optimize transcription accuracy. Automatic transcription was then performed using OpenAI's Whisper speech recognition system (medium model), which has demonstrated strong performance in child-directed speech contexts (Radford et al., 2023). All transcriptions were subsequently reviewed by trained researchers, who cross-checked audio and video recordings to verify content and assign speaker labels. For AI Chatbot-child and AI Chatbot-parent-child interactions, transcripts were directly extracted from the AI Chatbot interface and manually aligned with audio-visual records to ensure accuracy.

To quantify verbal engagement, trained coders identified conversational turns within each interaction transcript. A conversational turn was defined as a pairwise exchange consisting of a child utterance followed by a response from the conversational partner, or vice versa. Then the total number of the conversational turns were divided by the total duration of each session to control for the variations in the duration across sessions and participants. This operationalization of CTC ratio was applied uniformly across all conditions.

### 2.6. *fNIRS Data acquisition and processing*

Cortical hemodynamic responses were recorded using the NIRx NIRScout functional near-infrared spectroscopy (fNIRS) system. Data acquisition employed a standard prefrontal cortex (PFC) montage supplied by NIRx, which comprised 15 optodes—8 sources and 7 detectors—arranged with an approximate inter-optode distance of 3 cm and configured via a NIRSCap to have 40 channels [20 oxygenated hemoglobin (HbO]) and 20 deoxygenated hemoglobin (HbR) channels] (Figure 3a). The fNIRS technique relies on the transmission of near-infrared light through fiber-optic cables; source optodes emitted light at dual wavelengths (760 nm for HbO and 850 nm for HbR). This light penetrates the cortical surface and is subsequently captured by detectors, enabling quantification of hemoglobin concentration changes. Prior to task onset, the cap was secured approximately one inch above the brow line, and a signal quality check was conducted to verify optimal calibration.

Individual-level fNIRS analysis was conducted using the MNE package (v1.8.0) across all 40 channels. The data were first preprocessed at the individual level using wavelet denoising combined with a moving average filter (Piazza et al., 2020). Quality checks were then performed using MNE's scalp coupling index (SCI) on a channel-wise basis, with channels excluded if their SCI score were still below 0.5 after preprocessing. Consistent with other infant fNIRS literature, participants with more than 50% of channels registering poor-quality SCI would have been fully excluded from the study (Arredondo et al., 2022; Pinti et al., 2024; Siddiqui et al., 2022; Vahidi et al., 2024). However, no participant exceeded this threshold. The channel quality metric used for subject and channel exclusion was calculated on the combination of all runs after preprocessing. Two participants had limited usable data (less than 5.3 minutes per session) and were therefore removed from the analyses. For participants remaining after channel quality exclusion, we only included those who successfully completed at least two of the three conditions in the final analysis (n = 3).

Following quality assessment, raw optical density data were converted into hemoglobin concentration values using the modified Beer-Lambert law, in accordance with MNE's default pipeline. Subsequently, a first-level general linear model (GLM) was applied using the MNE statistics module. From this analysis, beta values for each condition (AI Chatbot alone, Parent alone, AI Chatbot and Parent Combined) were extracted and carried forward to a second-level analysis. The baseline used for the first-level GLM was the average activation across all three conditions. This was done by dropping the intercept in the design matrix prior to deriving the beta values (Tak & Ye, 2014). Beta values that exceeded three standard deviations from the mean were considered outliers and removed from the analysis. We linearly rescaled the Beta estimates so that each condition is expressed in micro-units. This transformation improves numerical interpretability while preserving all effect sizes and p-values.

Next, we defined six a priori prefrontal regions of interest (ROIs), dorsomedial PFC (DMPFC), dorsolateral PFC (DLPFC), and ventromedial PFC (VMPFC), each bilaterally (Figure 3b). Channel-to-ROI assignments were as follows: Left VMPFC (S2_D3, S3_D3, S5_D3, S5_D4), Left DMPFC (S3_D2, S3_D4, S4_D4, S4_D2), Left DLPFC (S2_D1, S1_D1, S1_D2), Right VMPFC (S5_D4, S5_D6, S6_D6, S8_D6), Right DMPFC (S4_D4, S4_D5, S6_D5, S6_D4), and Right DLPFC (S7_D5, S7_D7, S8_D7). For each ROI, we averaged activation across constituent channels to derive a single ROI-level estimate per participant. Due to variability in channel availability/quality (ranged from N=13-18), the effective sample size per ROI ranged from N = 16 to 18.

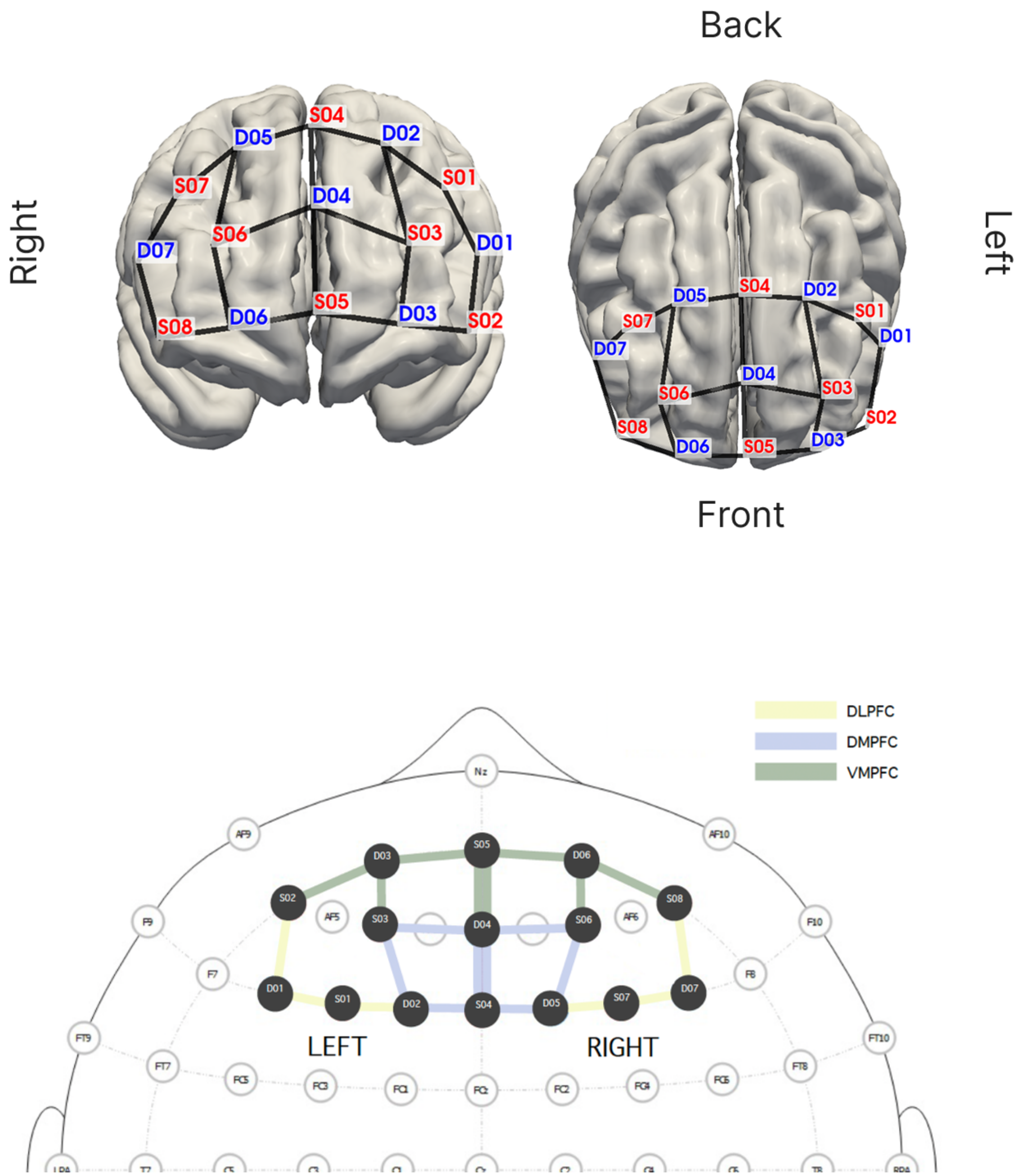

**Figure 3**. (a) NIRx prefrontal 8x8 montage 3D anterior view; (b) Six-ROI channel map. Channels are color-coded by a priori bilateral ROIs: vmPFC (ventromedial), dmPFC (dorsomedial), and dlPFC (dorsolateral).

*2.7.Analysis*

All analyses were conducted in SPSS 28.0.0.0 (IBM Corp.). Unless noted, tests were two-tailed with $\alpha = .05$. Greenhouse–Geisser corrections were applied where sphericity was violated; adjusted dfs are reported in the Results. For follow-up pairwise tests, familywise error was controlled with Bonferroni adjustment.

### *2.7.1 Behavioral and survey data.*

We first examined zero-order bivariate correlations among child demographics (age, sex), parent demographics (sex, educational levels, relationship status), and key study measures (anthropomorphism subscales, preference items, and behavioral indices). To evaluate anthropomorphism ratings, we fit a 4 (Domain: perceptive, emotive, imaginative, epistemic; within-subjects) × 2 (Actor: AI, parent; within-subjects) repeated-measures ANOVA. Significant effects were probed with Bonferroni-corrected comparisons on estimated marginal means; planned parent-vs-AI paired t tests were conducted within each domain.

Children's self-reported preferences were analyzed using one-way MANOVAs to test whether the set of chatbot anthropomorphism subscale scores differed by (a) "liked-most" session (AI-only, Parent-only, AI+Parent) and (b) "next-time" preference. For behavioral engagement, we analyzed the child conversational turn-contribution (CTC) ratio separately in AI-only and AI+Parent sessions. For each session, a MANOVA tested the multivariate association between CTC and the set of chatbot subdomain scores (Perceptive, Emotive, Imaginative, Epistemic). Pillai's Trace was used as the omnibus statistic given the small sample and potential covariance heterogeneity.

### *2.7.2 fNIRS data.*

ROI definitions followed the preregistered mapping, averaging channels within dmPFC and vmPFC bilaterally to obtain ROI-level HbO estimates per condition. Primary second-level models were repeated-measures GLMs with Condition (AI-only, Parent-only, AI+Parent; within-subjects) and anthropomorphism subscale scores entered in separate models as continuous between-subject predictors. For each significant ROI, we conducted sensitivity analyses that controlled for child age and sex. Multiple comparisons across ROIs and condition-wise tests were controlled using the Benjamini–Hochberg false discovery rate (FDR) at $q < .05$ (Benjamini & Hochberg, 1995). We report oxygenated hemoglobin (HbO) outcomes because HbO typically exhibits stronger associations with BOLD signals in fMRI and greater sensitivity to task-induced neural activity, and higher signal-to-noise than deoxygenated hemoglobin (HbR) in fNIRS(Gagnon et al., 2012; Huppert et al., 2006; Issa et al., 2016; Kinder et al., 2022). HbR results are provided in the Supplement material.

## 3. Results

### *3.1.Participant characteristics*

Participant demographics and key variables are summarized in Table 1. Zero-order correlations (see Supplementary Table S1) indicated that child age and sex were not associated with anthropomorphism scores or with behavioral/self-report engagement measures. Parent demographic characteristics were also not significantly related to these outcomes.

**Table 1**. Demographic characteristics and key variables

| Variable | Level | N | n (%) | Mean | SD | Min | Max |
|---|---|---|---|---|---|---|---|
| Child age (months) | | 23 | | 70.13 | 6.64 | 60 | 80 |
| Child sex | | 23 | | | | | |
| | Female | | 12 (52.2%) | | | | |
| | Male | | 11 (47.8%) | | | | |
| Child ethnicity (Hispanic/Latino) | | 23 | 2 (8.6%) | | | | |
| Child race (White/Caucasian) | | 23 | 23 (100.0%) | | | | |
| Parent sex | | 23 | | | | | |
| | Female | | 19 (82.6%) | | | | |
| | Male | | 4 (17.4%) | | | | |
| Parent ethnicity (Hispanic/Latino) | | 23 | 1 (4.3%) | | | | |
| Parent race (White/Caucasian) | | 23 | 23 (100.0%) | | | | |
| Parent education | | 23 | | | | | |
| | High school diploma | | 1 (4.3%) | | | | |
| | Associates | | 2 (8.7%) | | | | |
| | Bachelor’s | | 9 (39.1%) | | | | |
| | Master’s or higher | | 10 (43.5%) | | | | |
| | Other | | 1 (4.3%) | | | | |
| Parent relationship status | | 23 | | | | | |
| | Single | | 1 (4.3%) | | | | |
| | Married | | 22 (95.7%) | | | | |
| Child interview - Favorite session | | 19 | | | | | |
| | AI | | 8 (42.1%) | | | | |
| | Parent | | 4 (21.1%) | | | | |
| | AI+Parent | | 7 (36.8%) | | | | |
| Child interview – Favorite partner for future storytelling | | 20 | | | | | |
| | AI | | 6 (30.0%) | | | | |
| | Parent | | 3 (15.0%) | | | | |
| | AI+Parent | | 10 (50.0%) | | | | |
| | I don't know | | 1 (5.0%) | | | | |
| Conversation turns count (CTC) ratio | | | | | | | |
| AI only session | | 20 | | 0.026 | 0.011 | 0.0100 | 0.0533 |
| Parent only session | | 20 | | 0.116 | 0.035 | 0.0438 | 0.1730 |
| AI+Parent session | | 19 | | 0.022 | 0.009 | 0.0111 | 0.0400 |
| Child mood ratings | | 22 | | | | | |
| | Sad | | | 1.27 | 0.77 | 1 | 4 |
| | Scared | | | 1.27 | 0.70 | 1 | 4 |
| | Angry | | | 1.50 | 1.06 | 1 | 4 |

| | | | | | |
|---|---|---|---|---|---|
| Happy | | 3.64 | 0.73 | 2 | 4 |
| Anthropomorphism | | | | | |
| Perceptive – AI | 21 | 4.00 | 1.26 | 2 | 5 |
| Emotive – AI | 21 | 3.57 | 1.57 | 0 | 5 |
| Imaginative – AI | 21 | 3.71 | 1.27 | 1 | 5 |
| Epistemic – AI | 20 | 4.44 | 0.83 | 2 | 5 |
| Perceptive – Parent | 21 | 4.95 | 0.22 | 4 | 5 |
| Emotive – Parent | 21 | 4.86 | 0.48 | 3 | 5 |
| Imaginative – Parent | 21 | 4.57 | 0.75 | 2 | 5 |
| Epistemic – Parent | 20 | 4.85 | 0.37 | 4 | 5 |

**Note.** Ns vary by measure due to missing data.

### *3.2.Technology exposure*

Direct exposure to LLM chatbots was uncommon among children: 2/23 had prior AI-chatbot use. In contrast, 12/23 parents reported prior chatbot use. Smart-speaker use was more prevalent among children (12/21 used daily), most commonly Alexa, Siri, or Google Assistant; exposure to social robots/talking smart toys was rare. Most children used applications on smartphones/tablets weekly or daily. Because technology-use variables were highly skewed (e.g., only two children reported chatbot use), we did not test associations between these variables and anthropomorphism; cell counts were too small for stable inference. Table 2 reports technology-exposure descriptives. More details are also included in Supplementary Table S2.

**Table 2**. Parent and child technology experience and frequency of use (N = 23)

| Measure | Yes (%) | <Monthly % | Monthly % | Weekly % | Daily % | >1×/day % |
|---|---|---|---|---|---|---|
| Parent uses ChatGPT | 12 (52.2%) | 6 (50.0%) | 2 (16.7%) | 2 (16.7%) | 1 (8.3%) | 1 (8.3%) |
| Child uses ChatGPT | 2 (8.7%) | -- | 1 (50.0%) | 1 (50.0%) | -- | -- |
| Child uses smart speakers | 21 (91.3%) | 3 (14.3%) | 2 (9.5%) | 4 (19.0%) | 12 (57.1%) | -- |
| Child uses smart toys | 1 (4.3%) | -- | 1 (100.0%) | -- | -- | -- |
| Child uses smartphone apps | 19 (82.6%) | 1 (5.3%) | 1 (5.3%) | 15 (78.9%) | 2 (10.5%) | -- |
| Child uses tablet apps | 21 (91.3%) | 1 (5.0%) | 1 (5.0%) | 13 (60.0%) | 6 (30.0%) | -- |

*3.3.Anthropomorphism*

We tested differences in anthropomorphism across Domain (perceptive, emotive, imaginative, epistemic) and Actor (AI, parent). We conducted repeated-measures ANOVAs, applying Greenhouse–Geisser corrections where appropriate. The analysis revealed a main effect of Domain, $F(2.32, 44.03) = 4.20$, $p = .017$, $\eta^2p = .18$, and a main effect of Actor, $F(1, 19) = 14.00$, $p = .001$, $\eta^2p = 0.42$, with higher ratings for parent than AI overall. Bonferroni-adjusted comparisons on estimated marginal means showed Epistemic state ($M = 4.64$, $SE = 0.12$, 95% CI [4.40, 4.89]) was greater than Imaginative abilities ($M = 4.18$, $SE = 0.19$, 95% CI [3.77, 4.58]), $\Delta = 0.47$, $SE = 0.14$, 95% CI [0.06, 0.88], $p = 0.018$. All other domain contrasts were nonsignificant after adjustment.

The Domain × Actor interaction trended toward significance, $F(1.97, 37.47) = 2.95$, $p = 0.065$, $\eta^2p = .13$ (Figure 4). Post-hoc paired *t*-tests (parent vs. AI within domain) indicated higher parent scores for Perceptive abilities, $t(20) = 3.40$, $p = 0.003$, $d = 0.74$; Emotive state, $t(20) = 3.72$, $p = 0.001$, $d = 0.81$; Imaginative abilities, $t(20) = 3.29$, $p = .004$, $d = 0.72$; and Epistemic state, $t(19) = 2.41$, $p = 0.026$, $d = 0.54$ (Table 1). Although lower than parent scores, AI ratings were still relatively high, particularly for the perceptive abilities and epistemic state means exceeded 4 out of 5.

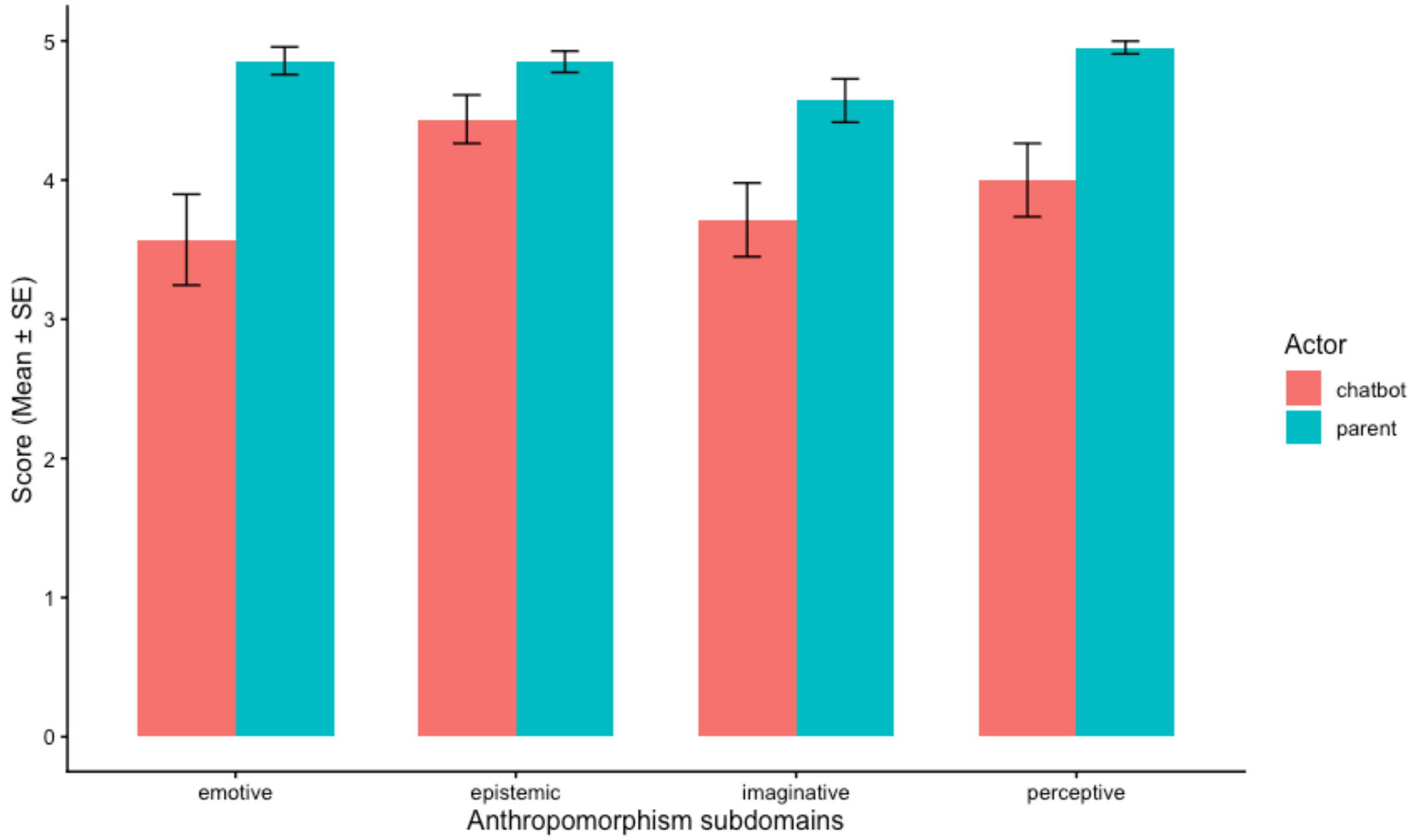

Figure 4. Bar graph of subdomains of child anthropomorphism for chatbot and parent .

*3.4.Self-reported preferences and behavioral engagement*

Children's favorite session was split between AI-only (8/19) and AI+Parent (7/19), with few preferring Parent-only (Table 1). A one-way MANOVA testing storytelling partner preference on the set of AI anthropomorphism subdomain scores was nonsignificant, Wilks' $\Lambda = 0.651$, $F(8, 24) = 0.72$, $p = 0.673$, $\eta^2 p = 0.19$. Preferences for the preferred conversation partner for a future interaction were similar. MANOVA showed no multivariate effect, Wilks' $\Lambda = 0.538$, $F(8, 24) = 1.09$, $p = 0.402$, $\eta^2 p = 0.27$.

During the AI-only session, the child CTC ratio was not associated with the set of AI anthropomorphism subdomain scores: Pillai's Trace = 0.269, $F(4, 13) = 1.19$, $p = 0.360$, $\eta^2 p = 0.27$. During the AI+Parent session, the multivariate test again showed no association, Pillai's Trace = 0.403, $F(4, 13) = 2.20$, $p = 0.126$, $\eta^2 p = 0.40$.

*3.5.Brain activation (fNIRS)*

We examined associations between anthropomorphism and HbO responses in left/right dmPFC and vmPFC. A repeated-measures model tested Condition (AI-only, Parent-only, AI+Parent; within-subject factor) and Perceptive abilities (between-subject factor), applying Greenhouse–Geisser corrections where appropriate. In right dmPFC**,** Condition was significant, $F(1.23, 17.20) = 7.53$, $p = 0.010$, $\eta^2 p = 0.35$, and the Condition × Perceptive abilities interaction was significant, $F(1.23, 17.20) = 10.47$, $p = 0.003$, $\eta^2p = 0.43$ (Figure 5). The interaction effects survived FDR correction for multiple comparisons at $q < 0.05$.

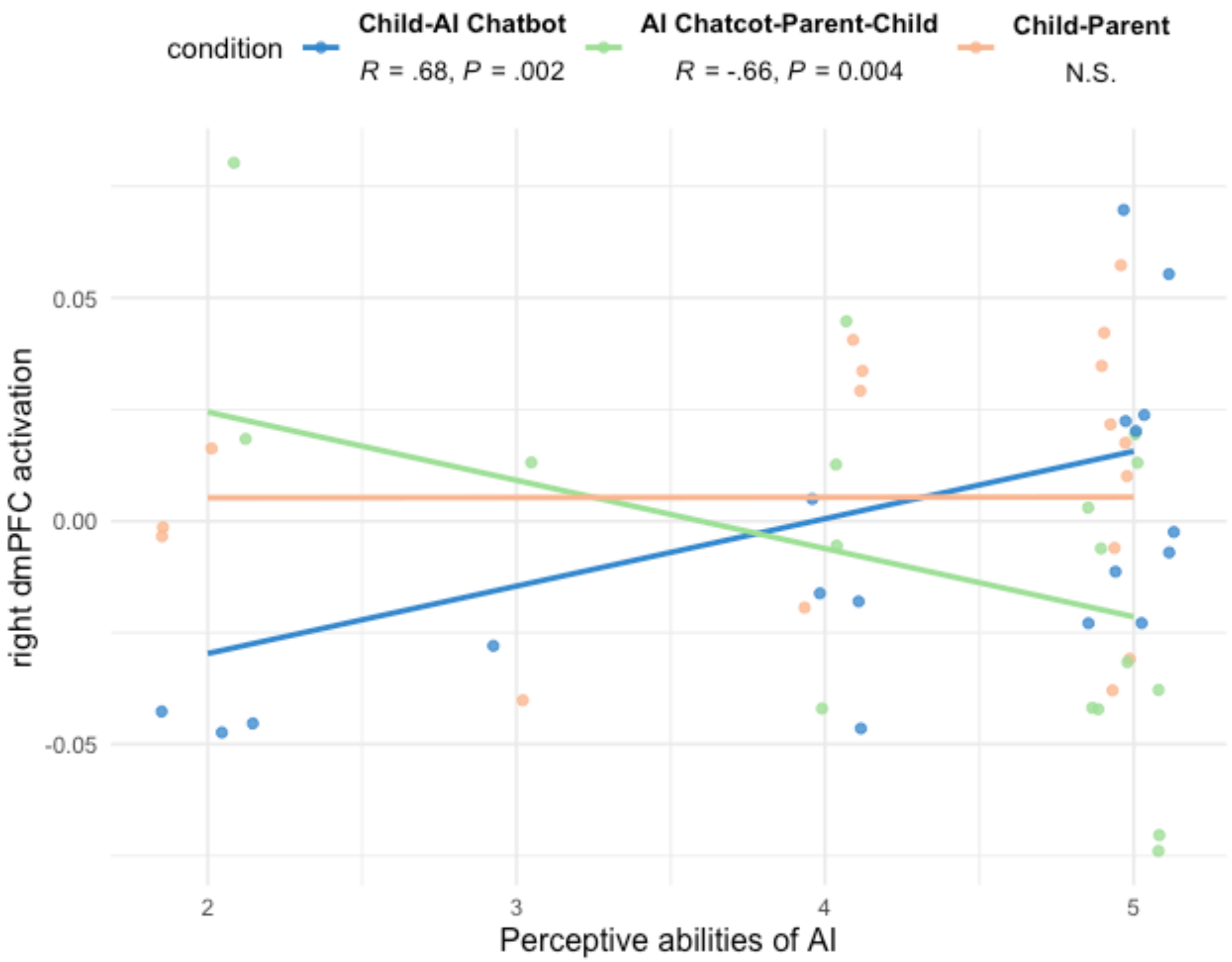

**Figure 5**. Scatter plot for child rated perceptive abilities of AI and right dmPFC activation during different task conditions. *N.S. = not significant.*

Post-hoc correlation tests showed that higher perceptive abilities were associated with greater right-dmPFC activation during the AI-only session and lower activation during the AI+Parent session (AI-only: $r(16) = 0.68$, $p = 0.002$; AI+Parent: $r(15) = -0.66$, $p = 0.004$). Parent-only correlations were nonsignificant ($p > 0.98$).

To address potential confounds, we repeated the right dmPFC model including child age and sex as covariates. The Condition × Perceptive interaction remained significant, $F(1.18, 14.17) = 9.04$, $p = 0.007$, $\eta^2p = 0.43$. Interactions of Condition with sex and with age were nonsignificant.

Main effects of sex and age were nonsignificant. Thus, results were consistent when controlling for these covariates.

Moreover, exploratory Pearson correlations examined associations between end-of-study mood ratings and right dmPFC activation by condition. Scared ratings were positively associated with AI-only activation, $r(16) = 0.586$, $p = 0.011$ (Figure 6). No other mood–activation correlations reached significance (Parent-only: $ps \geq .107$; all other associations: $ps \geq .070$). These exploratory results are descriptive and uncorrected for multiple comparisons.

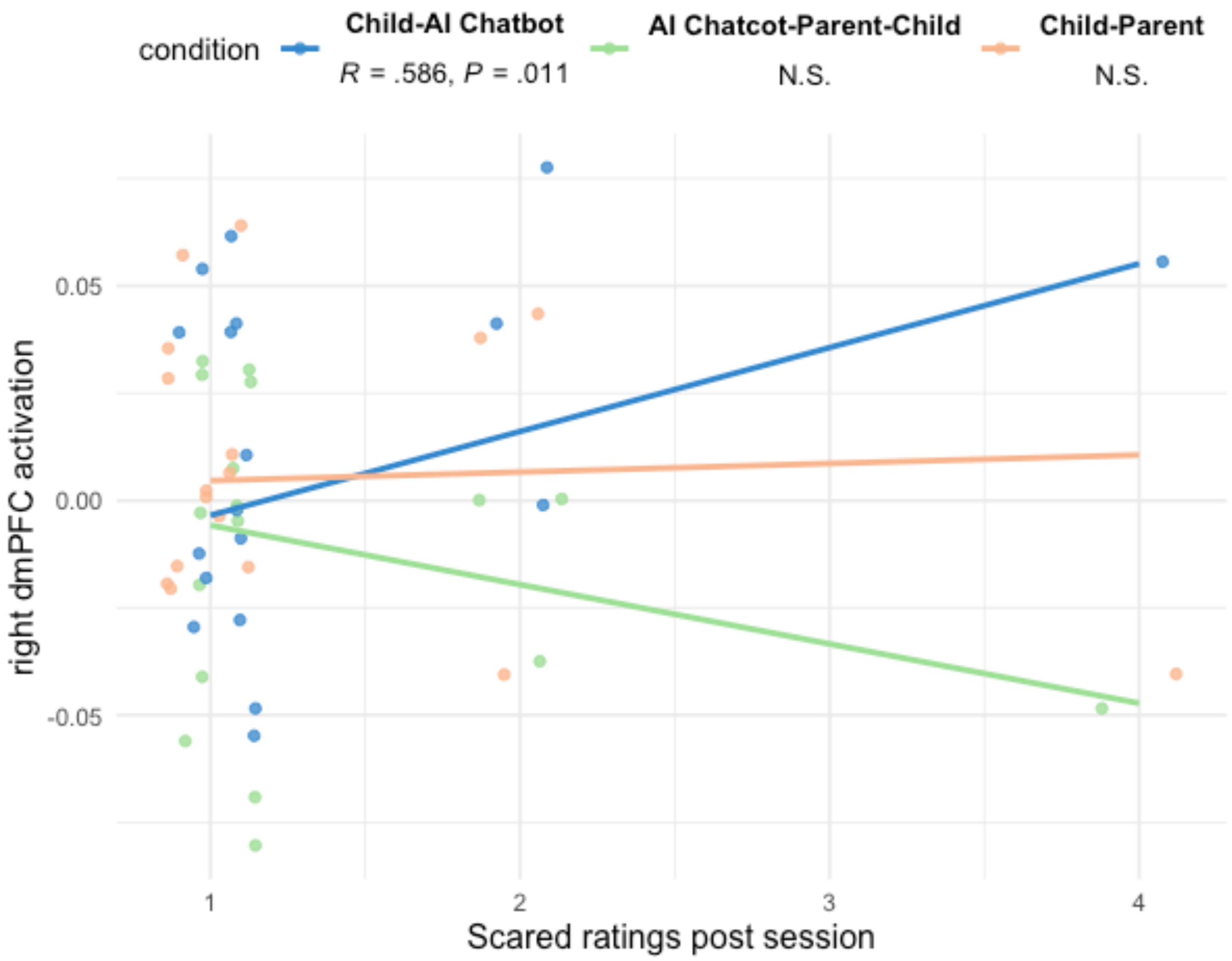

**Figure 6**. Scatter plot for child post session scared ratings and right dmPFC activation during different task conditions. *N.S. = not significant.*

## 4. Discussion

Young children are increasingly coming into contact with AI chatbots powered by LLMs through learning apps and entertainment platforms, yet we still know very little about how they understand these systems or how their perceptions translate into behavior during interaction. In this study, we focused on an important gap in the literature: children's tendency to anthropomorphize the AI chatbot, that is, the extent to which they think of the chatbot as having humanlike abilities such as seeing, hearing, knowing, feeling, or deciding. We measured anthropomorphism after children engaged in collaborative storytelling with the AI, with a parent, and with both together, and we asked how these beliefs related to children's behavior and to their brain responses. The inclusion of the parent conditions allowed us to compare the AI to a familiar human partner and to examine the role of a parent in the child's experience with the AI.

The findings suggest that children attributed humanlike qualities to the AI across perceptive, emotive, imaginative, and epistemic domains, although, as expected, they rated their parents higher overall. This is consistent with prior work showing that children at this age typically ascribe richer mental states to humans than to robots or other artificial agents (Di Dio et al., 2020; Manzi et al., 2020). At the same time, two domains stood out for the AI: children gave relatively high ratings on the epistemic items and on the perceptive items. By distinguishing these domains, our results suggest that children may be particularly sensitive to cues that signal the chatbot's ability to take in information provided by a child (perceptive) and to learn it, then create new information for storytelling (epistemic). The observed pattern is consistent with the dynamics of the interaction. The chatbot generated full stories and asked follow-up questions, which makes it look knowledgeable and capable of learning. It also responded with a face or gaze animation, which makes it look like it is perceiving. These are the cues that make machines seem more social and more mindlike to children in prior child–robot research (Kory-Westlund et al., 2017; Kory-Westlund & Breazeal, 2019b), and they align with classic explanations of anthropomorphism that emphasize human-likeness and uncertainty about how a system works (Epley et al., 2007).

Individual differences in anthropomorphic ratings were associated with the brain data. We focused on oxygenated hemoglobin with fNIRS over medial prefrontal regions and examined how children's anthropomorphism related to brain responses by condition. Children who strongly endorsed perceptive-ability items, such as thinking the AI can see or hear, showed higher activation in right dmPFC during the AI-only condition and lower activation in right dmPFC when interacting with the AI together with a parent. There was no reliable association in the parent-only condition. The effect size of the Condition × Perceptive Abilities interaction in right dmPFC was in the medium to large range and remained significant when we accounted for age and sex.

One possibility for why the dmPFC tracks perceptive anthropomorphism in this condition-specific way is that this region is selectively engaged when participants perceive an agent as having a mind, prompting mentalizing processes. The dmPFC is widely implicated in mentalizing and perspective taking, that is, thinking about what another agent knows, intends, or will do next (Healey & Grossman, 2018). When a child more strongly believes the chatbot can perceive, the child may try harder to infer the chatbot's intentions and to keep track of its point of view. That extra social inference may recruit dmPFC during AI-only interaction. By contrast, when a parent is present with the AI, the parent can help the child interpret ambiguous responses, fix errors, and explain what the AI can and cannot do. This social scaffolding can reduce uncertainty and the need for top-down inference, which might lower dmPFC activation. This social buffering by a parent is consistent with previous studies showing that parent presence can support children's engagement with media and help regulate children's stress in novel media contexts (Ewin et al., 2021; Stevens & Takeuchi, 2011; Wood et al., 2016).

The exploratory analysis further suggested a potential link between brain responses and affect in children. Children who showed higher dmPFC activation in the AI-only condition tended to report feeling more scared at the end of the overall session. Highly humanlike agents can sometimes feel eerie to children. As human-likeness increases and mindlike cues accumulate, discomfort can rise for some children, a pattern described as the uncanny valley (Brink et al., 2019; Cihodaru-Ștefanache & Podina, 2025). Even infants can show uncanny-valley responses to humanoid robot faces (Lewkowicz & Ghazanfar, 2012). In the context of the current study, strong anthropomorphism of perceptive abilities may be associated with heightened attention to the chatbot's social cues. This, in turn, may amplify children's cognitive and emotional load in their interactions with the AI, reflected in higher dmPFC recruitment and more uneasy feelings.

The findings suggest potential implications for design and practice. The overall pattern observed in the current study is consistent with the value of inviting parents into child–AI interaction, especially for young children who readily anthropomorphize. Design elements that encourage brief parent participation, small debrief moments about what the system can and cannot do, and clear signals about capability limits may help calibrate expectations without reducing engagement. These ideas align with ongoing discussions of age-appropriate AI literacy and with child-rights guidance emphasizing transparency to reduce anthropomorphism and adult mediation (Delaney & Chen, 2024; Rosman et al., 2025; Su, 2025).

Three points about the brain activation findings merit attention. First, although we controlled for age and sex in the dmPFC model and the interaction remained significant, prior exposure to voice assistants and tablets could also shape children's expectations about machine perception and memory. In our sample, the range of exposure to technology lacked sufficient variability to formally test this hypothesis. Including exposure as a covariate in larger samples will help rule out this alternative. Second, in our protocol, the AI + Parent session occurred last. Order and familiarity can themselves reduce uncertainty, which could partly explain lower dmPFC in that condition. A counterbalanced design in follow-up work will address this possibility directly. Third, the child's emotions were assessed after all sessions had ended. Thus, the ratings may reflect overall mood across sessions and may also be most affected by the final AI + Parent session. However, right dmPFC activation in the AI + Parent session was not directly associated with feeling scared, and the direction was negative at the trend level (Figure 5). Future studies should assess children's emotional ratings after each condition and relate them to condition-specific brain activation.

In the current study, anthropomorphism was not linked to a child's preference for the conversational partner or to our behavioral measure of engagement, which was the number of conversational turns. The null associations may reflect the limits of these particular measures. Which session a child liked most may compress a complex experience into a single forced choice. Similarly, because interaction styles differed between the AI and the parent, the turn-taking ratio had limitations in accurately reflecting child engagement with different partners. Future studies should examine finer-grained contingencies, such as how quickly partners respond, how often the child repairs misunderstandings, or whether the parent steps in to clarify what the AI meant, which may be more directly tied to social-cue processing and may show stronger links to anthropomorphism than a simple count of turns.

Several additional limitations are important to consider. First, our sample size provided power primarily for medium to large effects, so null behavioral associations should be treated cautiously. Second, we did not include a less humanlike AI comparison, which limits direct tests of human-likeness as a causal factor. Third, our fNIRS montage emphasized dmPFC and vmPFC regions; other brain regions involved in social cognition, such as the temporoparietal junction, will be important to examine in future studies. Fourth, the sample skewed toward White, highly educated families, which limits generalizability. Digital-divide disparities, where home broadband and computer access vary by income, parental education, and race or ethnicity among U.S. households with children, may shape opportunities to engage with and interpret AI technologies, potentially biasing observed patterns of anthropomorphism and engagement (Katz et al., 2017; Martin et al., 2024). Future studies should purposively recruit more socioeconomically and racially or ethnically diverse children and families, including those with limited broadband or device access.

In sum, we found that young children attribute substantial mental-state capacities to a social AI chatbot, with perceptive and epistemic attributions especially prominent. Conceptually, emphasizing perceptive and epistemic attributions clarifies which aspects of AI mind are most readily ascribed in early childhood and, in turn, which design features may most strongly shape children's expectations and interpretations of AI behavior. Perceptive abilities of anthropomorphism further relate to dmPFC engagement in a condition-specific way. Activation is higher during AI-only interaction and lower when parents are present, and higher AI-only activation aligns with more scared feelings at the end of the session. Practically, these findings underscore the value of transparent, developmentally appropriate signaling of system capabilities and limits, as well as the potential role of parent co-presence in scaffolding children's understanding of AI interactions.

## Funding sources

This work was supported by National Science Foundation (NSF) [2115008] and National Research Foundation of Korea [NRF-RS-2024-00340675].

## Declaration of competing interest

The authors declare that they have no known competing financial interests or personal relationships that could have appeared to influence the work reported in this paper.

## Acknowledgements

We thank Mohsena Ashraf, Seungwook Lee, Alina Mali, Elizabeth Wolfgram, and Matthew Zago for their assistant in study design, data collection, or data coding. We also thank the participants who contributed their time to this research.